\documentstyle[11pt,paspconf,epsf]{article}
\def\edcomment#1{\iffalse\marginpar{\raggedright\sl#1\/}\else\relax\fi}
\reversemarginpar

\begin{document}
\title{ Collapse and Outflow: Towards an Integrated Theory of Star Formation} 
\author{Ralph E. Pudritz$^{1,2}$, Dean E. McLaughlin$^2$, and
Rachid Ouyed$^2$}
\affil{$^1$CITA, University of Toronto, Toronto, Ontario, Canada M5S 1A1}
\affil{$^2$Dept. of Physics and Astronomy, McMaster University,   
Hamilton, Ontario, Canada L8S 4M1 }

\begin{abstract}

Observational advances over the last decade reveal that star formation is
associated with the simultaneous presence of gravitationally collapsing gas,
bipolar outflow, and an accretion disk.  Two theoretical views of star
formation suppose that either stellar mass is determined from the outset by
gravitational instability, or by the outflow which sweeps away the collapsing
envelope of initially singular density distributions.  Neither picture appears
to explain all of the facts.  This contribution examines some of the key
issues facing star formation theory.

\end{abstract}
\noindent To appear in {\it Computational Astrophysics: Proceedings of the
12$^{th}$ Kingston Meeting}, ed. D. A. Clarke and M. J. West (San Francisco:
ASP), in press
\section{Introduction}

How stars form is one of the most important unsolved problems in astrophysics.
It has turned out that the process is surprisingly rich, involving the
formation of dense cores in magnetized molecular clouds, gravitational
collapse, the ubiquitous presence of accretion disks around young stellar
objects, and most surprisingly perhaps, the presence of high speed bipolar
outflows and jets.  While gravitational collapse and the formation of disks
might have been expected in any model of star and planet formation, the role
of bipolar outflows has yet to be fully integrated into our thinking.

Star formation theory has scored some notable successes, among them being the
elucidation of the importance of magnetic fields, and the physics of
gravitational collapse and of magnetized outflows from the central engine.  In
spite of these advances however, there is still no generally accepted answer
to the most basic question of all; what determines the mass of a star?  In
this review, we discuss some of the main ideas of star formation with a view
to addressing this question.  Its solution will no doubt require sophisticated
mathematical and numerical tools.

\subsection{Basic Facts}

Star formation occurs within very specific, over dense regions within
molecular clouds known as molecular cloud cores (Benson \& Myers, 1989).  The physical
conditions within these cores presumably provide the initial conditions for
star formation.  One of the most basic properties of cores is that their mass
distribution is very well defined.  Measurements indicate that the clumps and
cores within molecular clouds obey a well defined relation in which the number
of cores per unit mass scales as $$dN(m)/ dm \propto m^{-\alpha},$$ where the
index $ \alpha = 1.6 \pm 0.2$ (e.g., Blitz 1991).

The internal structure of cores has been intensively studied in the last
decade.  Molecular cloud cores are known to be prolate structures (Myers et
al.~1991) in which rotation is insignificant in comparison with the
self-gravity of the core.  The line widths in molecular cloud cores {\it
increase} as one moves outwards to larger radii; thermal gas pressure can
dominate only in the innermost regions (.01-pc scales) in cores (Fuller \&
Myers 1992; Caselli \& Myers 1995).  Nonthermal motions dominate on larger
scales in cloud cores, with the nonthermal velocity dispersion in low mass
cores scaling as $$
\sigma_{NT} \propto r^{0.5}\ ,
$$ 
while for higher mass cores
$$
\sigma_{NT} \propto r^{0.25}\ . 
$$
These relations {\it appear to hold for both starless and star-containing
cores}.  We therefore appear to be seeing the initial conditions for star
formation.  Several theorists argue that this is misleading and that all cores
are already affected by outflows from newly formed stars.  If the nonthermal
line width in cores is produced by the interaction of bipolar outflows with
their surrounding core gas, then it has been argued that one could construct a
model for the stellar initial mass function or IMF (e.g., Silk 1995).

The source of energy that is sufficient to balance gravity within molecular
clouds and their cores is the magnetic field that threads them.  Zeeman
measurements show that molecular cloud and core fields have energy densities
comparable to gravity (Myers \& Goodman 1988, Heiles et al.~1993).  The
nonthermal line widths would presumably reflect a generic MHD turbulence, or
perhaps superposition of MHD waves in clouds.  No general theory for how such
MHD turbulence might be excited yet exists, although outflows have been
suggested as a possible source of excitation.
 
Of central importance for an integrated theory of star formation is an
understanding of why outflow and collapse are operative at the same time as a
star forms.  The so-called Class 0 sources, which are objects having virtually
no emission at wavelengths below 10 $\mu$m, and spectral energy
distributions characterized by single blackbodies at $T \simeq 15-30$ K, are
important in this regard.  There is some evidence to suggest that these are
protostars whose collapsing envelopes may exceed the central protostar in mass
suggesting an age of $t \simeq 2 \times 10^4$yrs in some models (e.g.,
Andr\'e, Ward-Thompson, \& Barsony 1993; but see Pudritz et al.~1996).  The
key point here is that such objects have particularly strong and well
collimated outflows (e.g., Bontemps et al.~1996).

Finally, infrared camera observations of embedded young stellar objects within
molecular cloud cores indicate that stars don't form individually, but as
members of groups and clusters.  Almost of necessity, the most detailed
available calculations of star formation focus on the formation of individual
stars.  However, this theoretical focus may be blinding us to the solution to
our basic question.

\subsection{Basic Ideas}

Theoretical thinking about the star formation process stems from two
fundamental, but quite different aspects of the physics of self-gravitating
gas clouds.  The first view is that stellar mass is determined by
gravitational instability, while the second is that it is determined by
shutting off accretion in collapsing cores.  There are persuasive arguments
for and against both of these pictures.

\medskip 
\noindent
{\bf Gravitational instability:} One of the classic calculations is Ebert
(1955) and Bonnor's (1956) analysis of the stability of an isothermal sphere
of self gravitating gas of mass $M$ that is embedded in an extermal medium
with a pressure $P_s$.  The critical mass for a cloud at temperature $T = 10$
K and supported purely by thermal gas pressure, is
$$
M_J = {1.2 (T/ 10\ {\hbox{K}})^2 \over (P_s/10^5 k_B\ {\hbox{cm}}^{-3}\ 
{\hbox{K}})^{1/2}}\ M_{\odot}\ .
$$
It is impressive that this argument picks out a solar mass so that one can say
that self-gravitating gas at 10 K naturally forms solar mass objects (e.g.,
Larson 1992).  The 10 K temperature arises from the balance of cosmic ray
heating and cloud cooling by millimetre radiation from collisionally excited
CO molecules (Goldsmith \& Langer 1978).   The Jeans mass
is significantly larger in turbulent media, where turbulent rather than purely thermal pressure
enters into the above expression (eg. Mckee et al. 1993).   

There are several problems with this view however.  There doesn't seem to be
an obvious way of explaining the initial mass function of stars, which ranges
over two decades in mass.  Why wouldn't the clump mass spectrum be the same as 
the IMF in this theory?  The measured mass spectral index
for the IMF is $\alpha_* = 2.35$ (Salpeter 1955).  Thus, while the total gas
mass in the CMF is dominated by its most massive core, the stellar mass in the
IMF is dominated by the low mass end.  This fact suggests that many low mass
objects prefer to form in the more massive clumps; i.e. cluster formation is
required (Patel \& Pudritz 1994).
Secondly, the role of outflows seems incidental to the process except insofar
as it removes the angular momentum of collapsed core gas allowing the star to
form via accretion through the disk.
 
\medskip
\noindent 
{\bf Truncating the collapse:} An equally fundamental view of a
self-gravitating cloud is that the accretion rate in the collapse of singular
isothermal spheres is fixed by molecular cloud core conditions (e.g., Shu
1977). For an isothermal equation of state, the accretion rate is a constant,
$$
\dot M = 0.975 {a_{eff}^3 \over G} = 1.0 \times 10^{-5} \left({a_{eff} \over
0.35\ {\hbox{km s}}^{-1}}\right)^3\ M_{\odot}\ {\hbox{yr}}^{-1}$$
where $a_{eff}$ is an effective sound speed in the core.  In these
self-similar theories, gravitational collapse and accretion onto a central
protostar can go on indefinitely.  Mass must therefore be fixed by the
mechanism that truncates the accretion phase such as jets and outflows.  
This view is interesting
because it incorporates outflow into the basic mechanism of star formation.

This view also has its problems.  As with the first theory, there seems to be
no obvious way in which an IMF could be produced.  
The role of the CMF is equally mysterious.  Secondly, while jets do pack
considerable power, they also appear to be highly collimated.  This is
especially true of outflows associated with the so-called class 0 sources.  An
outflow that doesn't cover a fair fraction of 4$\pi$ is unlikely to be able to
eject the remains of a collapsing envelope.  Such material could still end up
on the disk, and accrete from there onto the central star.

\section{Initial States}

The basis for the gravitational instability picture arises most simply in the
model worked out by Bonnor and Ebert.  Consider applying an external surface
pressure $P_s$ on an isothermal cloud of mass $M$.  Calculate the structure of
the resulting clump that is in hydrostatic balance with its own gravity and
internal pressure $P$ and the external pressure.  For such pressure bounded
equilibria, we may ask the question, at what radius $R$ will one find pressure
balance between clump pressure and $P_s$?  This problem is solved by finding
solutions to the Lane-Emden equation for these truncated configurations.  If
one now asks for the characteristics for our isothermal, non-magnetic cloud,
that is {\it critically stable} ($dP_s/dR = 0$), Ebert and Bonnor found that
$$ M_{crit} = 1.18 {\sigma_{ave}^4 \over (G^3 P_s)^{1/2}}$$ 
$$ \Sigma_{crit} = 1.60 (P_s/ G)^{1/2}\ .$$
These models have a finite central density and attain a $\rho\propto r^{-2}$
structure at large radii.

The generalization of this analysis for arbitrary equations of state may be
found in McLaughlin \& Pudritz (1996), where one finds that the expressions
for $M_{crit}$ and $\Sigma_{crit}$ for clouds with any equation of state
will differ from the isothermal, Bonnor-Ebert ones at the $<10\%$ level; the
physical scalings remain the same. (The effect of the equation of state is
to modify the numerical value of the line width $\sigma_{ave}$; see McLaughlin
\& Pudritz 1996.) The gradual loss of magnetic flux by ambipolar diffusion
implies that cores are supported against gravitational collapse in their
innermost regions for about ten free fall times.  As long as the central
regions of cores are magnetically supported, their central densities continue
to grow slowly.  
Detailed calculations show that once a critical value of the
ratio of the gas mass to magnetic flux in the central region is
surpassed, then the central density profile of magnetized cores
begins to approach a singular solution more rapidly.  During this more dynamic phase, 
collapse probably begins before a singular state is actually achieved
(e.g., Basu \& Mouschovias 1994).  

Shu (1977) and Li \& Shu (1996) argue that this steepening of
the density profile continues until the density actually becomes singular.  In
this event, which occurs say at time $t = 0$, its structure is simply
described by the relation;
$$ \rho(r) = { \sigma^2 \over 2\pi G}\ r^{-2}\ .$$
This singular distribution has a finite mass at the centre (related to a
numerical constant, $m_o$).  Once it is achieved (in finite time), it is
impossible for gas in the vicinity of the newly-formed protostar to evade
gravitational collapse.  Thus the the main accretion phase in singular
isothermal sphere (SIS) models, consists of an ``inside-out'' collapse of the
envelope onto this protostar. The evolution during this phase is then best
discussed in terms of an outwards-moving collapse front --- the so-called
``expansion wave.''

\subsection{ Equations of State}

A theoretical model of star formation in more massive cores must incorporate
some means of describing the nonthermal motions.  One way of proceeding is to
model the gas with an effective equation of state (EOS).  If one resorts to
polytropic models as an example, then the total line width scales as $\sigma^2
\propto P / \rho \propto \rho^{\gamma - 1}$.  Now since the lines are observed
to broaden as one goes to larger physical scales (where the density is
decreasing), then this simple result requires models with $\gamma < 1$, which
brings up the idea of negative polytropic indices (Maloney 1988).  Lizano \&
Shu (1989) modeled the general structure of cores by breaking the pressure up
into isothermal and turbulent contributions; $P = P_{iso} + P_{turb}$.  In
order to handle the turbulent motion, they suggested a so-called logotropic
relation (their words) between turbulent gas pressure and density; $ P_{turb}
= \kappa\ \ln(\rho/ \rho_{ref}$). On the other hand, McKee \& Zweibel (1995;
see also Pudritz 1990) later noted that a gas dominated by the pressure of
Alfv\'en waves would have an effective EOS of the form $P_{wave}\propto
\rho^{1/2}$.

\begin{figure}
\plottwo{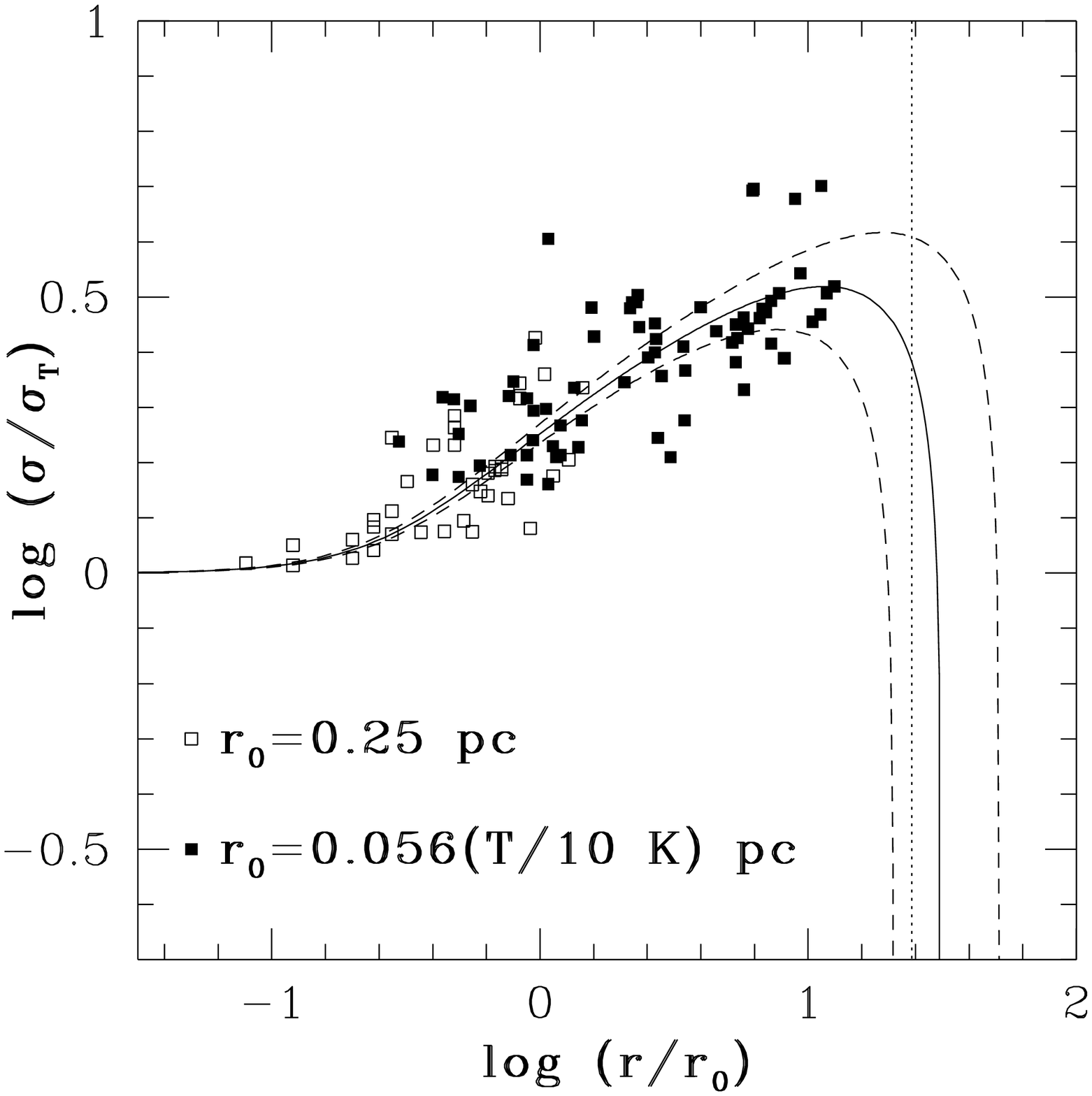}{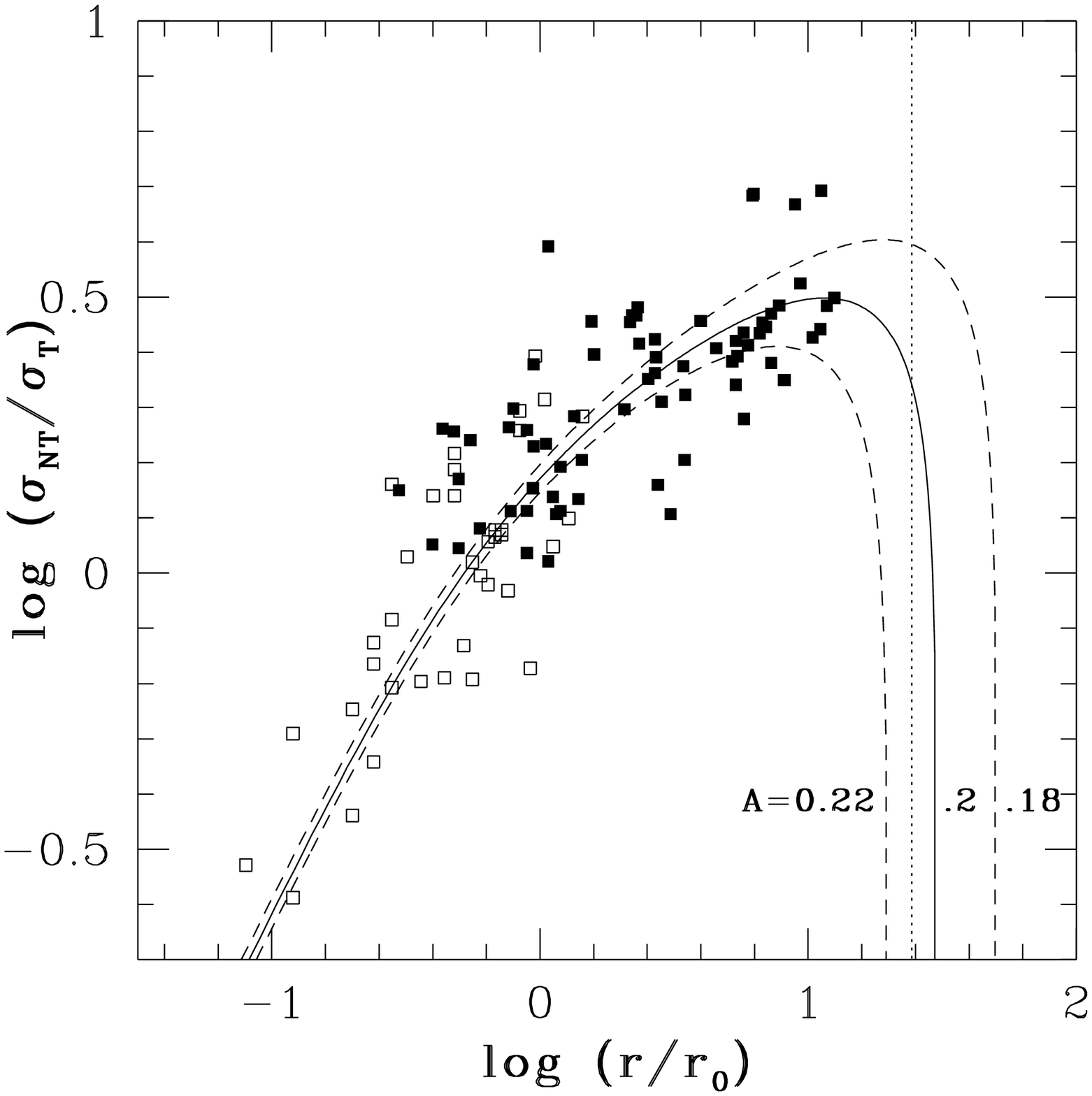}
\caption{Total line width $\sigma$, and turbulent component
$\sigma_{\rm{NT}}$, relative to the thermal contribution $\sigma_{\rm{T}}$.
Data on low mass cores ({\it open symbols}) are referred to a larger scale
radius than high mass cores ({\it filled symbols}); the former are less
centrally concentrated.  Curves are the predictions of the logotropic EOS, for
three values of the parameter $A$; vertical lines mark the radius of a
critical-mass, $A=0.2$ logotrope (from McLaughlin \& Pudritz 1996).}
\end{figure}

The data of Caselli \& Myers (1995) provides a way of testing possible EOS.
The challenge is to fit the trends in both the low and higher mass cores using
a single EOS.  McLaughlin \& Pudritz (1996) found that the models mentioned
above did not fit the data.  Their best fit is achieved by the so-called pure
logotrope,
$$P_{total}/P_c = 1 + A\ \ln(\rho/ \rho_c)$$
$$ A \simeq 0.2\ ,$$
in which the {\it total} gas pressure $P_{iso}+P_{turb}$ has a logarithmic
dependence on density.  In the singular limit, these models have density
profiles that are much shallower than SIS models;
$$\rho(r) = (A P_c/ 2 \pi G)^{1/2}\ r^{-1}\ .$$

Figure 1 shows the fit of the pure logotrope to observed line widths inside
cores, and illustrates how well constrained the value of the coefficient $A$
is. The vertical lines mark the radius of the critically stable model.
Unmagnetized logotropes have critical masses of $92 M_{\odot}$ while
magnetized ones have critical masses of $250 M_{\odot}$. Such cores are
obviously far more massive than $1 M_{\odot}$, and must therefore be the
objects in which multiple star formation occurs. Since solar mass clumps fall
far below the critical mass for a logotrope, their internal structures are
dominated more by gas pressure than by gravity. Thus, they are less centrally
concentrated than higher mass clumps, and this is reflected in their larger
scale radius $r_o$ in Figure 1. Note that
star formation appears to be occurring in a wide variety of these clumps which
suggests perhaps that the gravitational instability analysis may have less to
do with the issue of stellar mass determination than does the accretion
picture.

\section{ Gravitational Collapse}

Insight into gravitational collapse in molecular cores has been gained by
considering special cases that are analytically tractable such as the collapse
of SIS models.  Non self-similar models, such as the Bonnor-Ebert solutions,
require a detailed numerical solution.  Thus, Foster \& Chevalier (1993)
investigated the collapse of Bonnor-Ebert spheres and found that their
numerical solutions produced supersonic velocities during initial stages of
the collapse.  Central inflow speeds reached -3.3 times sound speed, and a
central density distribution $\rho \propto r^{-2}$ developed.  Their work
generalizes the collapse calculations of Larson (1969) and Penston (1969) who
began with uniform spheres.  The difference in the results is that Foster \&
Chevalier find that supersonic speeds develop only in a small region in the
centre, and not throughout the model.  Also, the mass accretion rate is
constant only if the initial configurations are very highly centrally
concentrated.

Following Larson (1969) and Shu (1977), it is convenient to define similarity
variables; if $a_t$ (in general, time dependent) is the sound speed, then the
self-similar dimensionless length is 
$ x  = r / a_t t .  $
The equations of motion then imply an accretion rate
$$ \dot M (0,t) \propto a_t^3 / G\ .$$  
In what follows, we take the extreme cases of the SIS (Shu 1977) and the
singular, pure logotrope models (McLaughlin \& Pudritz 1997).  The different
character of the self-similar collapse solutions for these two different
models well illustrates the effect that the EOS has upon the physics of the
collapse.  For the SIS model, the position of the expansion wave (see \S1.2)
in the self-similar variable $x$ is always located at $x_{exp} = 1$.  The
expansion wave moves outwards through the undisturbed envelope at constant
speed $a_t=a$; its position at any time is then
$r_{exp} = a t\ ,$  
and the mass of the central protostar grows as 
$$M  = m_o a^3 t / G \qquad m_o = 0.975\ .$$ 

For the logotrope on the other hand, the expansion wave is located at $
x_{exp} = 1 / 4 \sqrt 2 $ and the sound speed is no longer a constant; $ a_t =
[A P_c 4 \pi G]^{1/2}t \propto t$. The position of the expansion wave in
space is 
$ r_{exp} = a_t t/4\sqrt{2} \propto t^2 $
and the mass of the protostar grows as
$$ M = m_o [A P_c 4 \pi G]^{3/2} t^4 / G \qquad m_o = 0.667\times10^{-3}\ .$$
The infall speed at any time is lower for the logotrope than for the SIS model
because the latter has a more centrally concentrated density profile.  Note
also that the numerical constant $m_o$ which scales the initial protostellar
mass at the instant $t=0$ is much smaller for the logotrope.  This has the
consequence that it takes much longer to grow low mass protostars.  

One of the
main results of McLaughlin \& Pudritz (1997) is that the time required to
accumulate a solar mass star is of order $2 \times 10^6$ years, which is much
longer than for an SIS model.  On the other hand, all stars in the logotrope
picture accumulate in the roughly the same time which is not true of SIS
models.  This has a major impact on our ideas of IMF formation.  It suggests
that star formation in the logotropic picture, must really be sequential in
time.  Indeed, any theory for the formation of star clusters 
must guarantee that low mass star formation gets
started first, since when massive stars turn up, the molecular cores will be obliterated.
While SIS models could only pertain to low mass cores, high mass
star formation necessarily takes place in more turbulent conditions so that
star formation
time scales are much shorter  
(eg. Myers \& Fuller, 1992).  Thus, here too, 
sequential star formation 
needs to be invoked.

\section{ Outflows}

Episodic jets are observed in AGNs, regions of star formation (e.g., Edwards
et al.~1993), and binary systems with black holes. Whenever one observes a
jet, there is good evidence that an accretion disk is also present; a fact
that is probably not fortuitous.  Young stellar objects have associated
outflows that last a long time, at least $ 1 - 2 \times 10^5$ yrs according to
Parker et al.~(1991).  The outflows in Class 0 submm sources have mechanical
luminosities that rival the accretion luminosity of the central object with
$L_{mech} \simeq L_{bol}$.  In all outflows, radiation pressure fails by
several orders of magnitude to provide the observed thrusts in winds so that
mechanisms involving magnetic drives seem to be suggested.

Current models for outflows invoke magnetic fields that thread Keplerian
disks.  They are of two types; (i) {\it hydromagnetic disk winds} wherein the
engine consists of a Keplerian disk threaded by a magnetic field that is
either generated in situ, or advected in from larger scales (e.g., Blandford
\& Payne 1982; Camenzid 1987; Lovelace et al.~1987; Heyvaerts \& Norman 1989;
Pelletier \& Pudritz 1992; Li 1995; Appl \& Camenzid 1993; K\"onigl \&
Ruden 1993); or (ii) {\it X winds}, which are magnetized stellar winds where
the interaction of a protostar's magnetosphere with a surrounding disk results
in the opening of some of the magnetospheric field lines (Shu et al.~1987,
1994).

Perhaps the most important difference between these two classes of models lies
in the role of the central object.  For disk winds, only the depth of the
gravitational well created by the central object is of any importance.  The
energy source for the flow is the gravitational energy release of material in
the Kepler disk as the wind torque extracts its angular momentum.  This view
implies that the physics of jets from the environs of protostars, or black
holes is essentially the same.  For X wind models on the other hand, the
magnetization and structure of the central object is critical.  Its magnetic
field strength must be sufficient to carve a magnetosphere inside the disk and
outflow requires that the magnetopause and co-rotation radii of the star are
virtually identical; $R_m \simeq R_{co}$.

These two different wind mechanisms make different predictions about the
possibility of truncating the collapse of the surrounding envelope.  The
X-wind model has a low density, radial component to the wind that could
possibly clear out the envelope.  Disk wind simulations, such as those of
Ouyed, Pudritz, \& Stone (1997) and Ouyed \& Pudritz (1997 a,b) (see below)
find that a finite fraction of the disk is involved in outflow and that
outflows are rapidly collimated towards the outflow axis.  This implies that
they may not be able to clear out the envelope.  As far as we are aware,
extensive calculations of this type have never been done in either of these
theoretical models so the jury is still out.

Numerical simulations of disk winds by Ouyed et al. (1997; see also Ouyed, 1977) 
were run in order to test
the predictions of steady state theory and to see whether or not
time-dependent calculations would yield jets that are truly episodic.  The
simulations have an initial state consisting of a central point mass, the
surface of a surrounding Keplerian accretion disk (inner radius $r_i$), and a
disk corona that is in exact (analytical and numerical) hydrostatic balance in
the gravitational field of the central object and in pressure balance with the
accretion disk below.  The disk and corona is threaded by a magnetic field
configuration chosen to have initial current ${\bf J} = 0$ so that no magnetic
force is exerted initially upon the corona.  Two different magnetic
configurations were investigated; the first was a vacuum solution for a field
with a conducting plate at its base (called a potential distribution), and the
second was a constant uniform magnetic field that is parallel to the z-axis
and perpendicular to the disk.  This second configuration was chosen because
no outflow is expected in steady state theory.  The models depend on 5
parameters; three prescribe the initial corona (ratio of gas to magnetic
pressure, thermal to rotational energy density, and the ratio of the density
of the base of the corona to disk density; all these measured at $r_i$), one
gives the ratio of the toroidal to poloidal field strength in the disk, and a
final parameter measures the speed at which mass is injected from the disk
into the base of the corona.  This latter speed is taken to be a thousandth of
the local Kepler speed, or a hundredth of the disk sound speed.  All lengths
in our simulations are in units of $r_i$, and all times ($ \tau$) are in units of the
Kepler time $t_i = r_i/v_{K,i}$ at the inner edge of the disk.

Simulations using the potential field configuration (see Ouyed et al.~1997,
Ouyed 1977)
clearly show a bow shock that separates the outflow that has started from the disk, with
the undisturbed corona.  The field lines and flow behind the bow
shock are collimated towards the $z$ axis into a jet-like, cylindrical
outflow.  
The result shows that a cylindrically
collimated, stationary outflow is achieved.  Cylindrical collimation is
predicted to be a generic feature of magnetized outflows in which the dominant
toroidal field of the outflow together with its associated current (which
flows up the jet) together exert a pinching Lorentz force towards the outflow
axis (e.g., Heyvaerts \& Norman 1989).  Ouyed et al. also show the position of
the Alfv\'en and fast magnetosonic (FM) surfaces where the outflow speed
achieves the propagation speeds of two of the three important wave speeds in
magnetized gas.  The data are compared with the position of
the Alfv\'en point on each field line in the simulation as predicted
by steady state theory (e.g., Blandford \& Payne 1982).  The agreement is very
good.  This and many other diagnostics (see Ouyed et al.~1997) show that there
is good agreement with steady state disk wind theory and our simulations.
While this is important and interesting, nature prefers to produce highly
time-dependent, episodic outflows.  Why is this?

Figure 2 shows that outflow occurs even for our initially uniform magnetic
field configuration.  The highly collimated, jet-like outflow is in this case,
dominated by a series of dense knots that are produced periodically on a time
scale of $t_{knot} \simeq 11 t_i$.  Knots are produced in a generating
region close to the central source; at a distance $z_{knot} \simeq 6 -7 r_i$.
Knots continue to be produced for as long as we have run our simulations, up
to 1000 time units, and so they are truly generic and are not transients.
Figure 2 shows three snapshots of a highly zoomed-in simulation ($( z \times
r) = (20 \times 10)r_i$) designed to show the details of the knot generating
region.  The left panels show the poloidal magnetic field structure of the
flow at three times during which a new knot is formed.  The right panels show
the opening angle of the magnetic field lines as a function of their footpoint
radius $r_o$ on the surface of the accretion disk.  One sees from these graphs
that the field lines have been pushed open in a small region of the disk,
making an angle of $50^\circ-60^\circ$ with respect to the disk surface.
These field lines have been opened up by the toroidal magnetic field pressure
arising from the Keplerian rotation of each field line.  Since Kepler rotation
is faster at smaller radii, one expects that torsional waves introduce
stronger toroidal field into the corona at smaller radii.  This creates the
radial gradient in toroidal field pressure that opens the field lines to less
than the critical angle of $60^o$ from the axis (Blandford \& Payne 1982).

\begin{figure}  
\plotfiddle{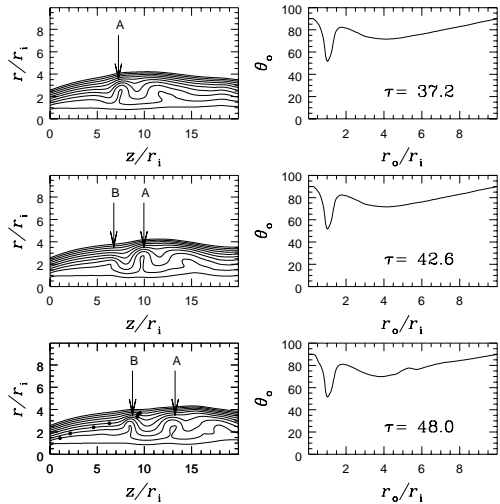}{2.4truein}{0}{35}{35}{-130}{-65}
\caption{The left panels show the magnetic field structure of the knot
generating region, at the three times; 37.2, 42.6, and 48.0 inner time units.
The right panels show the angle $\theta_{o}$ of field lines at the base of the
flow, with the disk surface, at these times.  Note the narrow band of field
lines which is sufficiently opened ($\theta_{o} \le 60^\circ$) so as to drive
the outflow. Only field lines involved in the knot generation process are
shown; field lines at larger disk radius stay reasonably vertical as seen in
the right panels (from Ouyed et al.~1997).}
\end{figure}

We found that knots are produced whenever the toroidal field in this inner
region is sufficient to recollimate the newly accelerated gas back towards the
outflow axis (see Ouyed et al.~1997).  The gas necessarily speeds up.  Because
the gas is rotating however, it encounters a centrifugal barrier at $r \ge
r_o$.  As it reflects off of this barrier, it collides with the slower gas
around it and shocks.  The shocked gas regions move away from this generator
region, and are kept coherent by strong enhancements of the toroidal field
both ahead of it, and behind it.  The knots, which are the overdense regions,
have low toroidal field strengths, and conversely, the space between the knots
is dominated by high toroidal field strength.  The time scale for the passage
of an Alfv\'en wave (in the toroidal field) from the jet radius towards the
axis and back again, turns out to be precisely the knot generation time.  This
episodic behaviour of jets may reflect on the nature of the accretion disk
that is feeding gas into the corona.  If the entry ram pressure of newly
injected material in the corona, exceeds the toroidal field at the base of the
corona, then we found that the outflow develops into a stationary flow.  Thus,
the general time-dependent behaviour of episodic jets may be intimately
related to conditions in the underlying accretion disks.  One must ultimately
remove the constraint of keeping the disk as a fixed boundary condition in the
problem if one hopes to explore this idea by numerical simulation.

\section {An Integrated Model?}

What general points about an integrated star formation theory arise from these
considerations?  Perhaps the least controversial point is that accretion disks
may be the glue that binds outflow and infall together.  Outflows may commence
as soon as the collapse has been sufficient to create even a tiny,
centrifugally supported region in a disk (e.g., Pudritz et al.~1996).
Accretion of infalling material onto and through the disk will drive the
outflow.  Thus, the continuous feeding of the disk by the infalling envelop
should help to sustain a vigorous outflow.  It is completely unclear as to
whether or not the details of the gravitational collapse are important for the
formation of an outflow (e.g., singular logrotropic vs.~isothermal collapse;
or non-self-similar collapse).  It is sobering to note that no self-consistent
numerical simulation of collapse that we are aware of has shown that
outflows are produced.  If outflows are disk winds, then their efficient
removal of disk angular momentum would help to drive an accretion flow through
the disk.  Of course, significant turbulent disk viscosity, such as could be
produced by MHD 
Balbus-Hawley (1991) turbulence, could also transport disk angular
momentum (radially).  The high collimation of hydromagnetic disk winds makes
it unlikely that they will clear out the infalling envelope.  X-winds, if they
occur, may have less of a problem in this regard.

While many details need to be checked before any useful predictions can be
made, we suggest from all of this that the physical processes of collapse and
outflow at the level of the formation of an individual star, have no obvious
means of dictating the mass of a star.  If this is correct, then the answer to
our basic question must take place in more general, larger scale processes.
Thus, the idea that stars form as members of groups and hence must somehow
compete for their gas supply, may be of central importance to the theory of
star formation.

\acknowledgments
REP thanks the organizers of this most stimulating conference for the
invitation to give this talk.  REP also acknowledges the financial support
and stimulating environment of CITA where he worked on this paper
during a sabbatical leave.   The research of DEM and RO was
supported by McMaster University, and that of REP through an operating grant
from the Natural Science and Engineering Research Council of Canada.

\end{document}